%% file: digital_twin_magazine_arXiv.tex
\documentclass[lettersize,journal]{IEEEtran}
\input{input.tex}
\usepackage{epstopdf}
\usepackage{enumerate}
\usepackage{algorithm}
\usepackage{amsmath}
\usepackage{float}
\usepackage{hyperref}
\usepackage{color}
\usepackage{makeidx}
\usepackage{bm}
\usepackage{cleveref}
\usepackage{url}
\usepackage{steinmetz}
\usepackage{varwidth}
\usepackage{soul}
\usepackage{xcolor,colortbl}
\usepackage{multirow}
\usepackage{balance}

\begin{document}

\title{Real-Time Digital Twins:  \\ Vision and Research Directions  for 6G and Beyond}

\author{Ahmed Alkhateeb, Shuaifeng Jiang, and Gouranga Charan \thanks{The authors are with the School of Electrical, Computer and Energy Engineering, Arizona State University, (Email: alkhateeb,  s.jiang,  gcharan@asu.edu).}}
\maketitle

\begin{abstract}
	 This article presents a vision where \textit{real-time} digital twins of the physical wireless environments are continuously updated using multi-modal sensing data from the distributed infrastructure and user devices, and are used to make communication and sensing decisions. This vision is mainly enabled by the advances in precise 3D maps, multi-modal sensing, ray-tracing computations, and machine/deep learning. This article details this vision, explains the different approaches for constructing and utilizing these real-time digital twins, discusses the applications and open problems, and presents a research platform that can be used to investigate various digital twin research directions. 
\end{abstract}

\section{Introduction} \label{sec:intro}

Heading towards 6G, communication systems are increasingly featuring key trends such as the employment of large numbers of antennas and the use of higher frequency bands \cite{Rappaport2019}. These technology trends bring higher data-rate and multiplexing gains to the networks, but also impose critical challenges on the ability of these systems to support highly-mobile, energy-constrained, reliable, and low-latency applications. For example, the deployment of large antenna arrays is associated with high channel acquisition and beam sweeping overhead \cite{Alkhateeb2014, lu2014overview}, which makes it hard for these massive MIMO systems to support mobile applications, and the use of higher frequency bands at mmWave and sub-THz makes the wireless links very sensitive to line-of-sight (LOS) blockages, which challenges the reliability and latency of the networks \cite{Andrews2016modeling}.

In this article, we present a novel vision in which a \textit{real-time} digital twin is utilized to make operational physical, access, network, and application layer decisions to the real physical world in communication and sensing systems. The key features of the envisioned digital twin-based wireless systems can be summarized as follows: 
\begin{itemize}
\item \textbf{Enabled by 3D maps and multi-modal sensing:} The envisioned digital twin will leverage precise 3D maps and fuse multi-modal sensory data from distributed devices and infrastructure nodes to construct an accurate real-time digital replica of the physical world. 

\item \textbf{Capable of making real-time decisions:} Leveraging advances in real-time 3D ray-tracing, efficient computing, and machine/deep learning, the envisioned digital twin will be capable of making real-time decisions for the wireless communication systems.

\item \textbf{Continuously refined for better approximation:}  We envision the digital twin as a model that will be continuously refined to improve its approximation of the physical world (including the electromagnetic and optical aspects) and to enhance its decision accuracy. 

\item \textbf{A global digital twin shared between devices:} In its ultimate version, we envision this digital twin to be global and shared between devices such that all devices can jointly enhance it and benefit from it using their coordinated sensing and communication decisions. 

\item \textbf{Used for all communication layer decisions:} With the real-time emulation of the physical world, the envisioned digital twin can be utilized to make physical-layer decisions such as channel prediction, as well as access, network, and application layer decisions. 
\end{itemize}

It is important to highlight here that the general concept of real-time digital twins has been studied before in the context of smart manufacturing, intelligent transportation, and healthcare \cite{wu2021digital}. For wireless communications, prior work has mainly focused on network operation topics such as edge computing, network optimization, and service management, in which digital twins are leveraged to simulate the real world at the network level \cite{khan2022digital,9374645}. In contrast, we focus on utilizing the real-time digital twin to simulate the real world with a particular emphasis on the physical modeling of the environment and wireless signal propagation. To that end, the envisioned real-time digital twins produce real-time instantaneous and statistical information about the wireless channels, which could be leveraged to make decisions for the physical, access, network, or application layers of the communication systems.

The goal of this article is to expose the potential of real-time digital twins for wireless communication and sensing systems in 6G and beyond. In the next sections, we discuss the key enabling technologies, the different approaches for constructing and utilizing these digital twins, and the various applications and future research directions. We also present a research platform that could be used to investigate many interesting digital twin research directions.

\begin{figure*}[t!]
	\centering
	\includegraphics[width=1\linewidth]{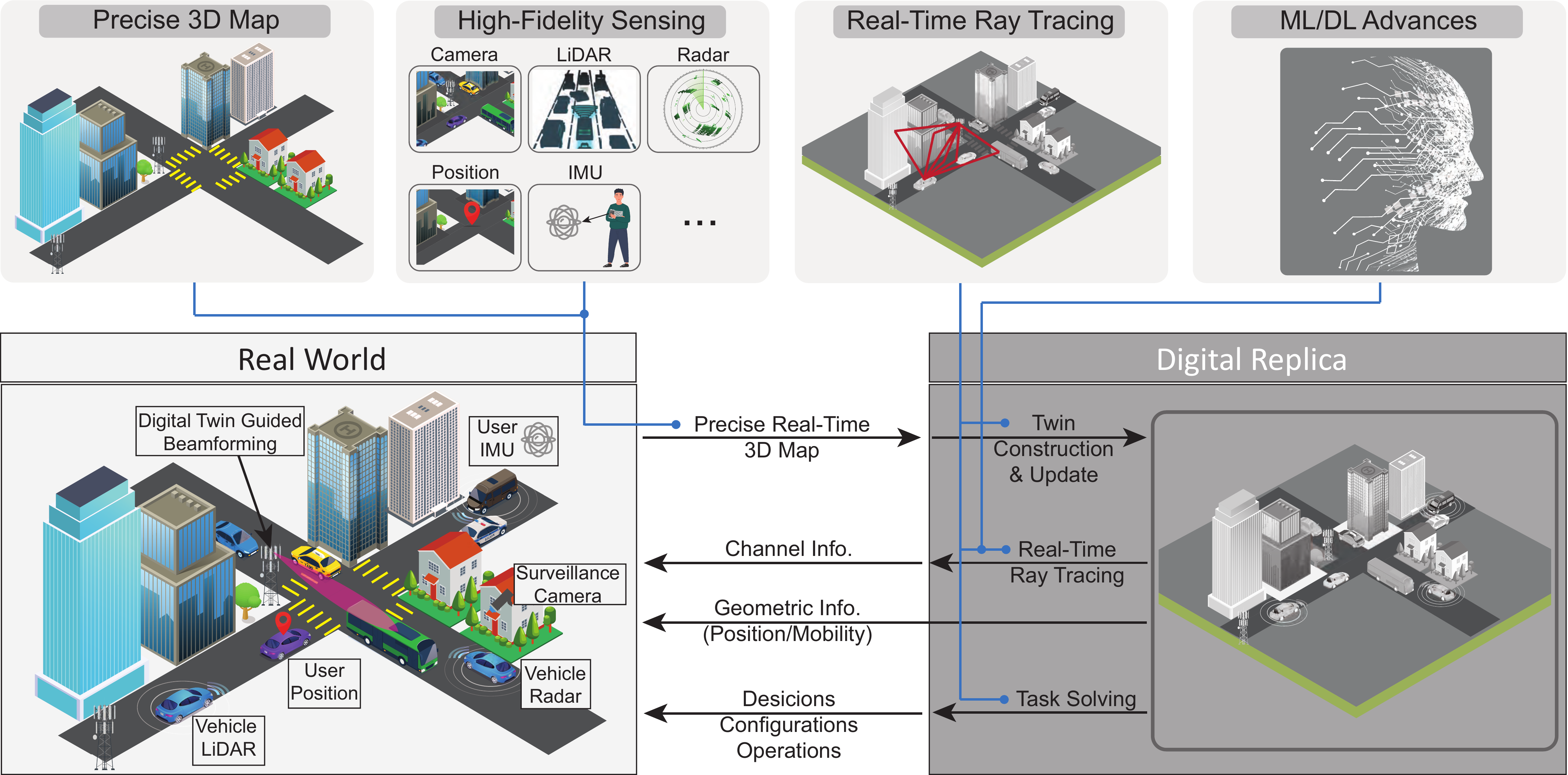}
	\caption{This figure presents the general idea of the digital twin with the four key enablers: precise 3D map, high-fidelity sensing, real-time ray tracing, and ML/DL. The digital twin is constructed based on a real-time 3D map. By performing real-time ray tracing, the digital twin infers channel information. The channel and geometry information can facilitate various applications.}
	\label{fig:overview}
\end{figure*}

\section{Today’s Technology Advances Lead to Real-Time Digital Twins} \label{sec:advances}
The vision of building real-time digital twins is motivated by the recent advances in 3D maps, multi-modal sensing, ray tracing, and machine/deep learning. Next, we briefly discuss these four key enablers for real-time digital twins.  

\textbf{Precise 3D Maps:} 3D maps contain information about the positions, shapes, orientations, and materials of the communication devices and other objects in the environment. The current application trends of AR/XR, autonomous driving, and metaverse technologies created an increasing demand for precise 3D map data. In response to that, the 3D map data collection, processing, and management capabilities have been significantly advanced. Vehicle, airborne, and satellite 3D imaginary sensors are now used to collect and build very accurate 3D maps. Further, the growing computational and database resources are making it possible to process and manage large-scale 3D maps, even at the scale of the full world like Nvidia OmniVerse \cite{NvidiaOmniverse}. Thanks to all these developments, precise 3D maps are becoming more affordable and accessible. 

\textbf{High-Fidelity Sensing:} Recent trends in sensing-aided communication, integrated sensing and communication (ISAC), and internet-of-things (IoT) tend to deploy multi-modal sensors, such as cameras, radars, LiDARs, positioning, at the infrastructure, user equipment, and IoT devices. These distributed sensors can complement each other since they have different observing angles and different types of information, such as position, shape, and mobility measures about the various stationary and moving objects. This sensing capability can be further enhanced by leveraging the recent advances in multi-modal data fusion \cite{liu2022multi}. As a result, it is becoming more feasible to acquire high-fidelity sensing information about the surrounding environment in nearly real-time, which is a key enabler for the envisioned digital twin.  

\textbf{Real-Time Ray Tracing:} Ray tracing simulators attempt to trace the wireless signal propagation paths between transmit and receive antennas, and generate the parameters, such as the angles of arrival/departure and complex path gains, of these propagation paths. A main limitation of the ray tracing simulators is that they typically require considerable computational overhead, and hence, incur high latency. The significant advances in parallel computing hardware and ray-tracing computational approaches over the last two decades, however, are enabling real-time ray-tracing for both wireless and optical signals. This means that, given precise real-time 3D maps, the wireless channels between the (possibly mobile) transmitters and receiver can potentially soon be computed in the digital twins in real-time.

\begin{figure*}[t!]
	\centering
	\includegraphics[width=1\linewidth]{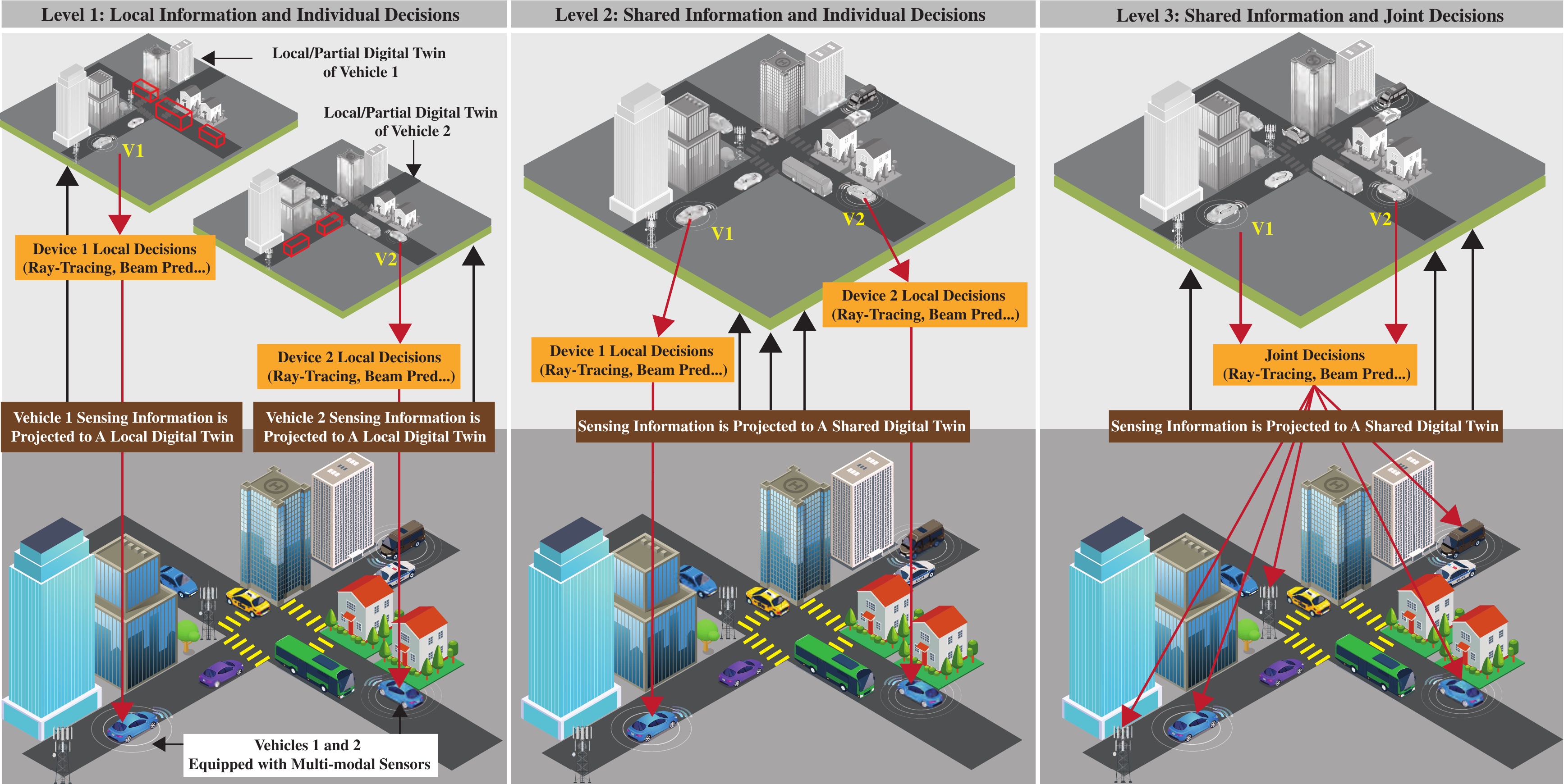}
	\caption{This figure presents our vision of how the digital twin system will operate in the real world. In particular, it shows three different operating modes (i) local information and individual decision, (ii) shared information and individual decision, and (iii) shared information and joint decision.    }
	\label{fig:three_levels}
\end{figure*}

\textbf{Advances in Machine/Deep Learning:} Machine and deep learning have demonstrated powerful capabilities in extracting features, approximating complex functions, and solving non-trivial optimization problems in wireless communication and sensing/perception. In the digital twins, these advances in machine/deep learning can be utilized to (i) enhance the quality and reduce the cost of building precise 3D maps, (ii) improve the efficiency of the multi-modal sensing in terms of sensory data processing, transferring, sharing, and fusion \cite{blasch2021machine}, and (iii) advance the ray tracing accuracy and reduce its latency and computational complexity.

All these advances in precise 3D maps, high-fidelity sensing, real-time ray tracing, and machine/deep learning are making it more feasible to realize real-time digital twins. In particular, the precise 3D map and high-fidelity sensing complement each other; while the 3D maps mainly contain information about the static objects in the environment, the high-fidelity sensing can augment these maps with information about the dynamic objects in real-time. Conducting ray tracing on these real-time 3D maps leads to real-time wireless digital twins. Further, the advances in machine/deep learning can be utilized to enhance the various aspects of these real-time digital twins. 
The real-time digital twins open opportunities for novel capabilities and applications in wireless communication and sensing systems. These digital twins can, for instance, be leveraged to compute channel or channel covariance information, predict LOS link blockages, proactively predict handover and traffic steering, and even predict application-specific caching requirements.

\section{True Digital Twins That Keep Learning} \label{sec:true}

There are different approaches of how such a real-time digital twin could be leveraged. In this section, we first discuss two main approaches where digital twins are either only used for training machine learning models or for making real-time decisions. Then, we present our vision for true digital twins that keep learning and improving over time. 

\textbf{Digital Twins for Training ML Models:}
With the precise 3D maps and accurate ray-tracing simulators, we can build high-fidelity and site-specific synthetic datasets. These datasets could be utilized to train machine learning models for the various wireless communication and sensing tasks. In particular, these synthetic datasets could be generated in large-scale and with high variance, which is hard and expensive to collect in the real world. The machine learning models that are trained on these site-specific synthetic datasets could then be deployed to make inference/decisions for the physical world. Further, these models could be refined using limited real-world datasets for better and more robust performance. This approach relaxes the latency requirements as the digital twins are not used for real-time decisions. The drawback, however, is that these ML models are not benefiting from the global real-time sensing and awareness that the real-time digital twins have, which may limit their performance. 

\textbf{Digital Twins for Real-Time Decisions:}
Another approach is to use these real-time digital twins to directly make real-time or near real-time decisions for the physical-world communication and sensing systems. For example, an FDD massive MIMO basestation can use the real-time digital twin to predict the downlink channel or, at least, the dominant subspace of this channel, which saves large channel training and feedback overhead. Mobile users may also use this digital twin, for instance, to predict if their LOS links are going to be blocked by a moving scatterer or whether they need to switch to other beams. This approach can, therefore, benefit from the real-time nature of the digital twin and the richer awareness about the surrounding environment in making real-time decisions. The drawback, however, is that the decisions that are solely made based on the digital twin will be very sensitive to the modeling accuracy, which challenges the robustness of these decisions.

\textbf{True Digital Twins:} The previous two approaches have a clear trade-off between the ability to benefit from the real-time awareness of the digital twins to make efficient decisions (e.g., in terms of wireless resources and mobility support) and the ability to ensure the robustness of these decisions. This motivates what we call \textit{true digital twins} that can be thought of as machine learning models themselves that keep learning and improving their approximation of the physical world, and hence their decisions, over time. In particular, these digital twins could leverage learning agents and use prior decisions and feedback to enhance the modeling accuracy of the 3D maps, the processing and integration of the multi-modal sensing data, and the approximation fidelity of the ray tracing. These true digital twins can, therefore, be used to make decisions that are both real-time and accurate for the various wireless communication and sensing tasks.

\section{Three Digital Twin Levels } \label{sec:levels}
Constructing and utilizing digital twins requires interaction with the devices that are contributing to the digital twins with sensing data and leveraging the digital twins in making decisions. This, however, raises questions about the required level of coordination between these devices for both the sensing and communication tasks. For that, we envision that digital twins will evolve through three main digital twin levels of coordination as the computation, synchronization, and communication capabilities develop over time. Next, we briefly present these operating modes (levels) and highlight how they could function in the real world.

\textbf{Local information and individual decision:}
We envision the first level of the digital twin to incorporate local sensing and individual decisions. In particular, each device equipped with one or more sensors will be expected to collect its local sensing information. The acquired local data can then be projected onto a 3D map to generate a real-time digital twin. This, in general, will help generate a partial view of the entire map, resulting in a localized digital twin. This local real-time digital twin can then be utilized to facilitate the sensing and wireless communication decision-making process. An advantage to such a local approach is that the limited sensing information captured by each device can be processed locally (on the device or edge), reducing the downstream processing and decision-making latency. However, the partial nature of the generated digital twin limits the scope of the decision-making capabilities of the devices. For example, enabling advanced applications, such as future blockage prediction and handoff, may require access to a more global real-time digital twin.

\textbf{Shared information and individual decision:} The proposed first level of the digital twin is limited by the local view of the wireless environment, which results in a partial digital twin. Generating an accurate real-time digital twin and reaping its benefits requires a global overview of the environment. Fusing the sensing information collected across different devices in the wireless environment at any given time can be a promising way to achieve this vision of a comprehensive digital twin. As such, this idea forms the basis of our vision of the second level of the digital twin. Similar to the first level, each device is expected to collect sensing data of its surrounding environment. However, here, we propose to fuse the information collected across devices to generate a detailed and thorough digital twin. The fusion operation can be performed at the edge or cloud. Even though information sharing across devices (or at the edge/cloud) is enabled, each device is envisioned to undertake its own sensing and communication decisions. This is due to the high computation, synchronization, and communication requirement for joint and globally-optimized decisions for all the devices.

\textbf{Shared information and joint/cooperative decision:} With the increased computation, synchronization, and communication capabilities of future devices, we envision that digital twins will evolve into a more global form where devices can also coordinate and jointly optimize their sensing and communication decisions. Therefore, in this third level of the digital twins, devices are expected to share and fuse their sensing information, mostly at the edge, similar to the second level, to form a global and comprehensive real-time digital twin. In addition, the sensing and communication decisions will be jointly optimized either in a central or distributed way. This is expected to enhance both the sensing and communication performance of future wireless systems. These three digital twin levels are illustrated in  \figref{fig:three_levels}, clarifying the differences in the information-sharing and decision-making approaches.

\begin{figure*}[t!]
	\centering
	\includegraphics[width=1\linewidth]{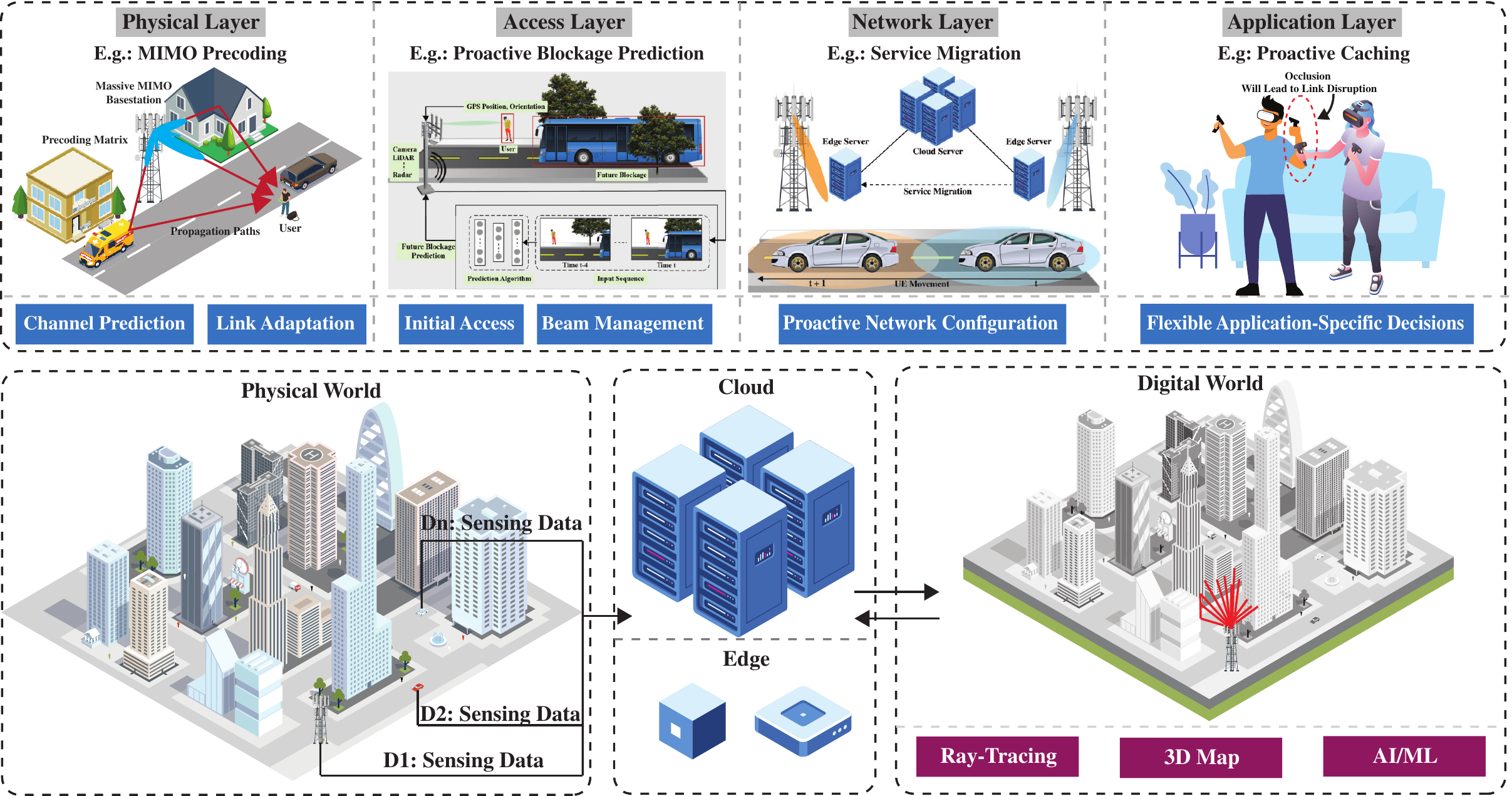}
	\caption{This figure presents some example communication applications that digital twins can enable across the different layers, such as the physical layer, MAC layer, network layer, and application layer. For example, in the MAC layer, the digital twin can proactively predict future blockages and initiate hand-off by tracking the mobility pattern of the different objects in the environment over time.}
	\label{fig:apps}
\end{figure*}

\section{Applications} \label{sec:apps}
The proposed digital twin leverages precise 3D maps and high-fidelity sensing to capture real-time information about the geometry and materials of the static and dynamic objects in the environment and to construct a real-time digital replica. By performing real-time ray tracing on this digital replica, we can infer various information about the communication channels, such as the propagation path parameters (path loss, delays, angles, etc.), channel and covariance information, link quality, and blockage status. Moreover, by utilizing the temporal  and spatial consistency of the channels, the digital twin can predict future channel information in dynamic environments. The real-time and future channel predictions can be leveraged to make real-time and proactive decisions that can potentially improve the communication system operations in the physical, access, network, and application layers.

\textbf{Physical Layer:}
The physical layer operation has, by definition,  a clear dependancy on the wireless channels. Many physical layer tasks, \textit{e.g.}, MIMO precoding and link adaptation, directly rely on partial or full knowledge about the communication channels. However, the channel acquisition is typically associated with high overhead, especially in large-dimensional systems, which degrades the overall system efficiency. 
Real-time digital twins open novel opportunities to revolutionize the channel acquisition process: When the real-time 3D maps and ray tracing computations are sufficiently accurate, the digital twin could be directly used to accurately infer the channels, reducing or even eliminating the channel acquisition overhead. Further, in scenarios where the approximation is not sufficient to accurately estimate the  full channels, the digital twin may still be used to predict the dominant channel subspaces and reduce the channel acquisition overhead \cite{sensing_otfs,Demirhan_mgazine_radar}. 
These digital twins can also be used to estimate the signal-to-noise ratio (SNR) of the communication links and improve the modulation and coding scheme selection. Another interesting physical layer application of the digital twin is to generate a massive amount of data for a given site in the real world. This site-specific data can then be utilized to optimize the  traffic beam sets and the channel feedback compression codebooks. 

\textbf{Access Layer:}
Current and future communication systems, especially at higher frequencies, employ large antenna arrays and highly-directional beams to achieve sufficient SNR and realize  high data rates. Aligning these beams, however,  typically requires large beam training overhead that scales with the number of antennas. The real-time and future channel information predicted by the digital twin can facilitate the initial access (initial beam alignment) and beam management for these systems and reduce the beam training overhead. Further, due to the increased penetration loss, high frequency communication links, e.g., in mmWave,  experience sudden disturbance due to blockages. By tracking the motion of the user and other objects in the environment, the digital twins can proactively predict the occurance and duration of incoming blockages before they happen. This enables seamless handover control and improve the network reliability and latency performance. Digital twins can alsp enhance the access layer resource allocation, user scheduling, MU-MIMO user pairing, and interference management, among many other applications.

\textbf{Network Layer:}
In prior work, digital twins have been used to optimize the network layer operations and applications such as device and traffic monitoring, resource allocation, edge computing, and cyber security. We refer to this type of digital twins as the network-level digital twins since they typically focus on monitoring network status and modeling the network-level entities, services, and dynamics. Differently, the envisioned physical-level digital twins can provide fine-grained information about the communication links, which can also be used to improve the accuracy of the radio access network (RAN) modeling. This accurate RAN  modeling can enhance the efficiency and reliability of various network-level operations. Further, the physical-level digital twins can provide real-time and future information about the user position and mobility characteristics, which could be leveraged to improve several edge computing operations. For instance, when a user is predicted to move out of the service area of the in-use edge server, proactive service migration can be triggered to improve the service quality. The physical-level digital twins can also work cooperatively and in an integrated manner with the network-level digital twins to enhance the end-user experience.

\textbf{Application Layer:}
New emerging applications, such as autonomous driving  and AR/VR, pose more stringent reliability and latency requirements to advanced communication systems. The 6G is envisioned to support 9-nine reliability with $0.1$ms latency for mission and safety-critical applications. Moreover, different applications often have diverse requirements for data throughput, reliability, and latency, thus different strategies when handling link instability. The digital twin can simulate the physical signal propagation and channels, which can offer fine-grained information on the communication link quality, both in real-time and proactively. This link quality information provides more flexibility for making efficient application-layer decisions in a way that respects this application diversity. For example, if a video streaming application knows ahead of time about the communication link disruption and blockage status, this application can pre-load a certain portion of the video, considering the duration of the cut-off, and achieve a seamless user experience with efficient usage of communication resources. While the digital twin provides low-level and fine-grained information about the communication links, how to efficiently utilize this information is still to be investigated.

\section{Digital Twin Research Platform} \label{sec:results}
Here, we present a digital twin research platform based on the DeepSense 6G \cite{DeepSense} and the DeepVerse 6G \cite{DeepVerse} datasets that can be used to investigate various digital twin research directions. Next, we briefly present these digital twin datasets and show how they can be used to investigate an example application of digital twin-assisted beam prediction. 

\textbf{Digital Twin Datasets:} The digital twins rely on co-existing real-world data and high-fidelity synthetic data generated using accurate 3D maps and ray tracing. For that, the DeepSense 6G and the DeepVerse 6G datasets are well-suited to facilitate the research, development, and application of digital twins. The DeepSense 6G is a large-scale multi-modal sensing and communication dataset collected in various real-world scenarios; the DeepVerse 6G is a synthetic dataset that can simulate high-fidelity multi-modal sensing and communication data from ray tracing. Combining real-world scenarios and their synthetic replicas from the two datasets, we present a digital twin research platform to investigate its efficacy, limitations, use cases, and applications in real-world systems.

\textbf{Example Application:} In \cite{digital_twin_conf}, the digital twin is used to first infer the channels with real-time 3D maps (containing user positions) and ray tracing. Then, these channels can be used to infer optimal beams. However, the high computational complexity of ray tracing may result in high latency, thus limiting real-time applications. For that, ML models can be employed to approximate the ray tracing simulation with accelerated computation. Furthermore, ML can be leveraged to learn the mapping function from 3D maps to the optimal beam in an end-to-end manner. This can potentially improve the computational efficiency since the ML has the flexibility to not model the unnecessary ray tracing for a specific communication problem. When the ML is trained solely using the digital replica, it is limited by the impairments in the 3D map and ray tracing. A small amount of real-world data can be utilized to fine-tune the ML models (\textit{i.e.} transfer learning) and to even transcend the digital replica.

\textbf{Experimental Results:} Scenario 1 of DeepSense is adopted as the real-world scenario, and the digital replica of this scenario is constructed and simulated. An ML model is first trained on the position and wireless beam data from the digital replica. Then it is fine-tuned on the real-world data. The top-2 beam prediction accuracy is obtained by testing the model on unseen real-world data. After training on 200 digital replica data points, the ML model can achieve a high top-2 accuracy of $91.4\%$. {Note that this digital twin approach does not need any real-world training data}. Moreover, with transfer learning, a small number of real-world data points (less than $20$) can quickly improve the ML model to go beyond the digital replica and achieve near-optimal performance. Next, the digital replica impairment is explicitly modeled by adopting a uniform beam steering codebook that is different from the beam codebook of the communication hardware. The uniform codebook leads to lower performance when fine-tuned on a very limited amount of real-world data. However, when more than 20 real-world data points are used for fine-tuning, the performance of the two codebooks becomes very similar. The mismatches between the real world and the digital replica can be efficiently calibrated by a small amount of real-world data.
\begin{figure}[t!]
	\centering
	\includegraphics[width=1\linewidth]{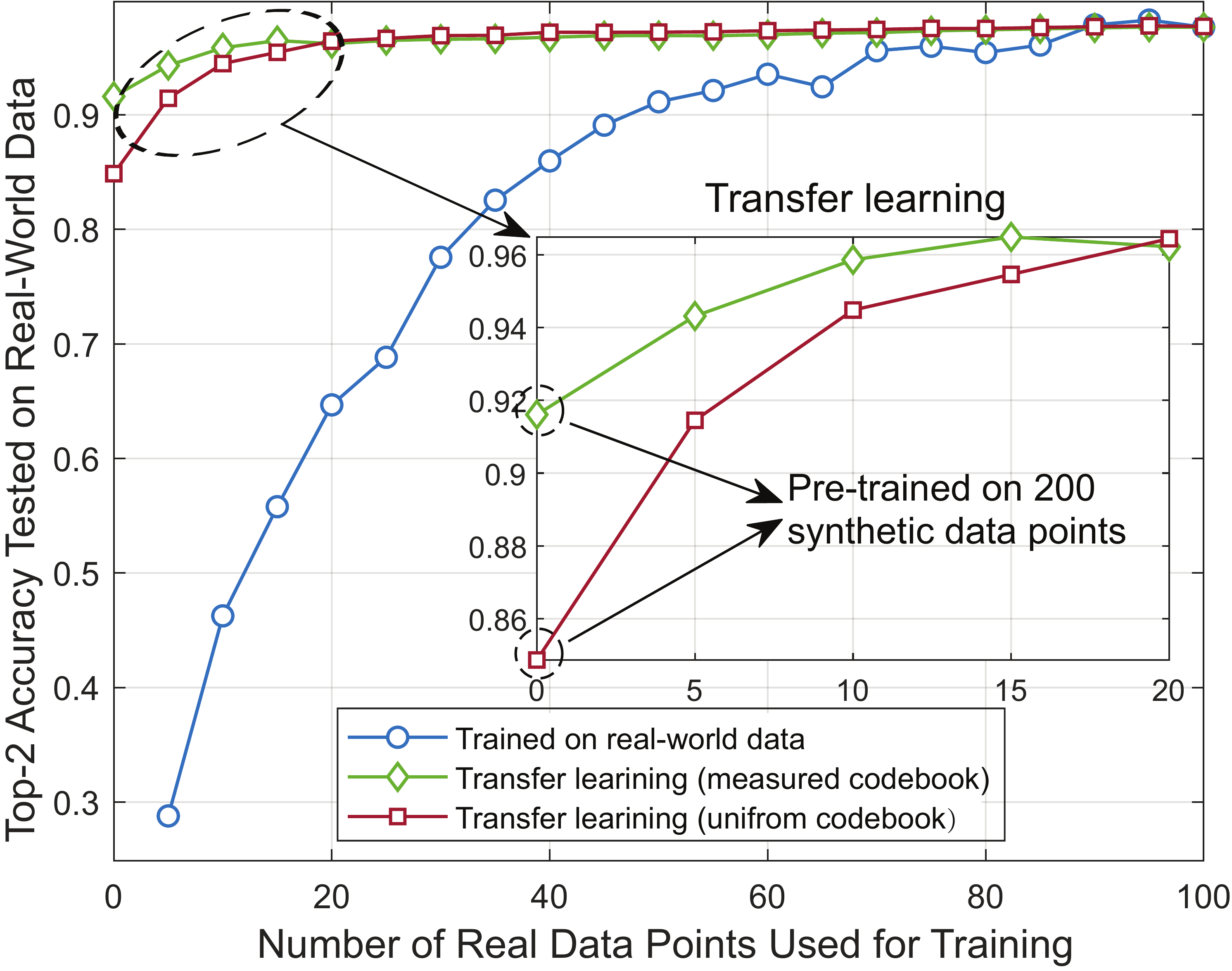}
	\caption{This figure shows the top-2 accuracy performance by first training the NN on the synthetic data generated by the digital twin and then fine-tuning it on real-world data.}
	\label{fig:transer_performance}
\end{figure}

\section{Future Research Directions} \label{sec:future}
Digital twins are envisioned to be an integral component of $6$G and beyond wireless communication systems and have recently piqued the interest of both the industry and academia alike. However, realizing this vision and exploring the true potential of digital twins requires overcoming the fundamental challenges and thoroughly investigating some of the important aspects. Next, we present some open research directions that need to be investigated toward enabling digital twin-aided next-generation wireless communication. 

\textbf{Impact of Digital Twin on Communication Tasks:} Deploying digital twin-aided solutions in the real world will require revisiting all the problem statements, such as LOS/NLOS beam prediction, blockage prediction, hand-off, etc., to evaluate the efficacy of these solutions accurately. In particular, it is necessary to investigate how much gain in performance can be achieved by utilizing the real-time digital twins compared to conventional and sensing-aided approaches. To that end, the recently published DeepVerse 6G synthetic dataset (with multi-modal sensing and communication data generated from 3D models and ray-tracing) was created to mimic the real-world scenarios of the DeepSense 6G dataset, effectively making them a digital twin of each other. These two datasets combined can enable the development and evaluation of digital twin-aided applications in real-world communication systems.

\textbf{Communication-Sensing Trade-Off:} In order to generate an accurate real-time digital twin with minimal latency, the sensing data collected across different devices, in some cases, need to be transferred quickly to a central unit for further processing (digital twin levels $2$ and $3$). The data transfer rate is dependent on the available bandwidth of the communication system itself. As the amount of sensing data increases (for example, with the increase in sensing modalities or the number of devices), so does the requirement for communication bandwidth. 
While access to diverse and detailed sensing information can help generate a more accurate digital twin, which improves the performance of the communication system, transferring that sensing data will consume communication resources, potentially offsetting any gains.
A promising solution to overcome this challenge is to first process the sensory data locally and extract relevant features, and then transfer the low-dimensional extracted features across devices. Investigating this trade-off and the possible solutions is an interesting open problem. 

\textbf{Sensing Fusion and Coordination:} As presented in Section~\ref{sec:levels}, in the third level of the digital twin, devices can communicate and coordinate to make joint sensing and communication decisions. For instance, multiple devices in the same location will capture similar sensing data. Most of this data may be redundant, and transferring this data over the limited communication bandwidth for further processing will only increase the computational cost overhead without significantly improving the generated digital twin. One solution to improve resource utilization can be to sense and transfer the optimum data required to generate a complete and accurate digital twin. Realizing such a solution will require the devices to communicate with each other and adopt efficient protocols for cooperative sensing. These challenges highlight the need for further investigation in this direction.    

\textbf{Central and Distributed Decision:} Generating an accurate and complete real-time digital twin necessitates the sharing and transferring of sensing data across devices. Given access to this digital twin of the real world, the next question is, how will this replica of the real world be utilized to facilitate the sensing and communication decision-making process? For instance, a user can make individual decisions in a distributed manner or adopt a centralized way that fosters a collaborative decision-making environment. Both these approaches have their advantages and limitations. For example, the individual distributed decisions are mostly made on the edge devices, which will minimize any latency involved with data transfer back and forth from the cloud devices. However, they generally require computation-intensive machine learning-based models, which will further increase the overhead in these resource-constrained edge devices. Developing a robust and efficient solution that can be deployed in the real world requires a detailed investigation of these different possibilities.

\section{Conclusion}
This article presented and discussed a vision for future wireless communication and sensing systems that would leverage precise 3D maps, distributed multi-modal sensing,  efficient ray-tracing computations, and advanced machine/deep learning to construct, update, and utilize real-time digital twins. As computations, communication, and synchronization capabilities evolve over time, we expect devices to gradually coordinate in building and updating global digital twins using their sensing information and potentially make joint sensing and communication decisions for the physical, access, network, and application layers. The article also presented a research platform, based on the real-world DeepSense 6G dataset and its digital replica, DeepVerse 6G, to investigate various digital twin research directions, and showed how to use this platform for one example application.

\balance

\end{document}

%% file: input.tex
\usepackage{amsfonts}
\usepackage{times}
\usepackage{graphicx}
\usepackage{latexsym}
\usepackage{dsfont}
\usepackage{amssymb}
\usepackage{amsmath}
\usepackage{cite}
\usepackage{verbatim}
\usepackage{subfigure}

\newcommand{\figref}[1]{{Fig.}~\ref{#1}}

\def\bb0{{\mathbb{0}}}

\def\bb{{\mathbf{b}}}

\def\b0{{\mathbf{0}}}

\def\sf0{{\mathsf{0}}}

